\documentclass[12pt]{iopart}

\begin{document}

\title{Vacuum self similar anisotropic cosmologies in $F(R)-$gravity}
\author{Pantelis S. Apostolopoulos$^{1}$}

\address{$^1$Technological Educational Institute of Ionian Islands, Department
of Environmental Technology, Island of Zakynthos, Greece\\
Emails: papost@teiion.gr; papost@phys.uoa.gr}

\begin{abstract}
The implications from the existence of a proper Homothetic Vector Field
(HVF) on the dynamics of vacuum anisotropic models in $F(R)$ gravitational
theory are studied. The fact that \emph{every} Spatially Homogeneous vacuum
model is equivalent, formally, with a ``flux'' -free anisotropic fluid model in standard gravity and
the induced power-law form of the functional $F(R)$ due to self-similarity
enable us to close the system of equations. We found some new exact
anisotropic solutions that arise as fixed points in the associated dynamical
system. The non-existence of Kasner-like (Bianchi type I) solutions in
proper $F(R)-$gravity (i.e. $R\neq 0$) strengthens the belief that curvature
corrections will prevent the shear influence into the past thus permitting
an isotropic singularity. We also discuss certain issues regarding the lack
of vacuum models of type III, IV, VII$_{h}$ in comparison with the
corresponding results in standard gravity.
\end{abstract}

\maketitle

\section{Introduction}

\setcounter{equation}{0}

\noindent Geometric symmetries have been used widely enough within the
context of General Relativity (GR) mainly because they lead to a significant
reduction of the complexity of the Field Equations (FE). In addition,
according to Noether's observations, the existence of special geometric
properties implies conservation laws and invariant quantities in the form of
first integrals \cite{Hall-Book, Duggal-Book}. The main disadvantage of
these approaches is the lack of sound physical motivation by assuming a
specific kind of geometric symmetry and the \textquotedblleft information
loss\textquotedblright\ from general models which leads to an incomplete
view of the whole picture. A counterexample, so far, to this situation is
the existence of a proper Homothetic Vector Field (HVF) admitted by the
underlying geometry of a large set of cosmological or astrophysical models.
Although the inspection of the role of homothetic (equivalently
self-similar) models has not be exhausted, their importance is well
established since transitively self-similar models represent the past and
future attractors for the majority of evolving (non) vacuum models \cite%
{Wainwright-Ellis, Ellis-Maartens-MacCallum, Coley-Book, Uggla:2003fp}.

\noindent Motivated from the above facts it is of mathematical and physical
interest to check if the above description of the asymptotic states of
general models also holds for the so called modified theories of gravity.
Considerable attention has been given to the extension of Einstein's theory
namely the quadratic theories of gravity in which curvature or/and Ricci
scalar invariants contribute in the Lagrangian. From a physical point of
view these theories appear to be an excellent enviroment to understand and
solve various problems in contemporary cosmology like the accelerated phase
of the Universe, the effect of quantum corrections to classical gravity or
the asymptotic isotropisation near the initial singularity or at future
times. The simplest version of the above alternative gravity theories is
represented by the presence of a functional $F(R)$ in the action integral
where $R$ is the Ricci scalar. A vast number of studies has been appeared so
far where the background geometry is described by the Robertson-Walker (RW)
metric. It has been argued that $F(R)-$theory could be a viable alternative
of standard gravity solving many open questions of cosmological interest
(see \cite{Sotiriou:2008rp, DeFelice:2010aj, Clifton:2011jh} for extensive
reviews and bibliography).

\noindent Because $F(R)-$theory is fourth order on the metric functions, the
non-linearity of the FEs and the coupling between temporal and spatial
dependence complexifies the analysis of the geometric and physical structure
of the specific model. Therefore it seems natural that few results exist in
the literature when we incorporate in $F(R)-$gravity more general geometries
than the RW. In cosmological scales the immediate departure from isotropy,
while holding the spatial homogeneity, of the Friedmann-Lema\^{\i}%
tre-Robertson-Walker (FLRW) universe is represented by the Spatially
Homogeneous (SH) or Bianchi geometries \cite{EinsteinSurvey,
Exact-Solutions-Book, Ryan-Shepley1}. Then the FEs are reduced to a coupled
system of ordinary differential equations that must be satisfied by the
anisotropy scale functions. The emergence of a power-law form of the metric
(equivalently the existence of a proper HVF) is justified since it
corresponds to fixed points of the associated dynamical system.

\noindent The details of the analysis described above can be found in the
present paper as follows: in section 2 we review some basic results
regarding the implications in the geometry and the dynamics of SH models
from the existence of a proper HVF. An important ingredient of our
discussion is the \emph{symmetry inheritance} property of the timelike unit
vector $u^{a}$ which characterizes the SH\ geometries. The vacuum $F(R)-$%
gravity within an anisotropic but homogeneous background is treated in
Section 3. Due to the fact that every SH vacuum model in $F(R)-$gravity is
equivalent with a \textquotedblleft flux\textquotedblright -free anisotropic
fluid model in standard gravity where the effective \textquotedblleft
dissipative\textquotedblright\ tensor is expressed in terms of the shear\ of
the comoving timelike congruence, the order of the resulting set of
differential equations is reduced to three. In section 4 we specialize our
study to vacuum self-similar models and derive the power-law form of the
functional $F(R)\sim R^{n}$ w.r.t. the Ricci scalar. Accordingly we give the
tetrad/expansion-normalized form of the effective FEs at the fixed points
which are used to determine a certain number of equilibria. We conclude our
analysis in section 5.

\noindent Throughout this paper we have used geometrized units such that $%
8\pi G=c=1$ and the standard index conventions: spatial frame and coordinate
indices are denoted by lower Greek letters $\alpha ,\beta ,...=1,2,3$, lower
Latin letters denote spacetime indices $a,b,...=0,1,2,3$.

\section{Spatial homogeneity and homothetic symmetry}

\noindent In cosmological setups the effects of the departure from the
standard FLRW model, are studied by using its simplest generalization, the
SH models. They are specified, in geometric terms, by requiring the
existence of a $G_{3}$ Lie group of Killing Vector Fields (KVFs) $\mathbf{X}%
_{\alpha }$ acting transitively on three-dimensional spacelike orbits $%
\mathcal{S}$. The conventional metric formalism \cite{EinsteinSurvey,
Exact-Solutions-Book, Ryan-Shepley1} then is used in order to express the
metric of the SH geometry in terms of the left-invariant 1-forms $\mathbf{%
\omega }^{\alpha }$ 
\begin{equation}
ds^{2}=-dt^{2}+g_{\alpha \beta }(t)\mathbf{\omega }_{a}^{\alpha }\mathbf{%
\omega }_{b}^{\beta }dx^{a}dx^{b}  \label{MetricSH-1}
\end{equation}%
where $g_{\alpha \beta }(t)$ are smooth functions of the time coordinate and
denote the spatial frame components of the induced three-dimensional metric,
constant in each spacelike hypersurface $t=$const.

\noindent In addition, there is a uniquely defined unit ($u^{a}u_{a}=-1$%
) timelike vector field $u^{a}$ normal to the spatial foliations $\mathcal{S}
$ \cite{Ellis:1998ct, vanElst:1996dr}:

\begin{equation}
u_{[a;b]}=0=u_{a;b}u^{b}\Leftrightarrow \frac{1}{2}\mathcal{L}_{\mathbf{u}%
}g_{ab}=u_{a;b}=\sigma _{ab}+\frac{\theta }{3}h_{ab}  \label{Veloc-Parts}
\end{equation}%
where $\sigma _{ab},\theta $ are the shear and expansion rates, associated
with $u^{a}$, according to the standard 1+3 decomposition of an arbitrary
timelike congruence and $h_{ab}=u_{a}u_{b}+g_{ab}$ is the projection tensor
normal to $u^{a}$ and represents the orthogonal metric of the instantaneous
rest spaces of the timelike observers (with an obvious abuse of notation we
can write $h_{ab}=g_{\alpha \beta }(t)\mathbf{\omega }_{a}^{\alpha }\mathbf{%
\omega }_{b}^{\beta }$). Because $u^{a}$ is irrotational ($\omega ^{a}=0$)
and geodesic ($\dot{u}^{a}\equiv u_{\hspace{0.2cm};b}^{a}u^{b}=0$), there 
exists a time function $t(x^{a})$ such that $u^{a}=\delta _{t}^{a}$ i.e.
each value of $t$ essentially represents the hypersurfaces $\mathcal{S}$.

\noindent On the other hand, one wishes to simplify further the geometry and
the dynamics of the SH models by assuming the existence of a Homothetic
Vector Field (HVF) $\mathbf{H}$ which is defined as 
\begin{equation}
\mathcal{L}_{\mathbf{H}}g_{ab}=2\psi g_{ab}  \label{HVFEquation1}
\end{equation}%
where $\psi =$const. essentially represents the (equal) time amplification
and space dilation.

\noindent The major gain from such a simplification \textquotedblleft
assumption\textquotedblright\ is not only the reduction of the FEs to a
system of algebraic equations but the fact that any solution of this system
represent the asymptotic state of evolving SH (or even less symmetric)
models. In the coordinates adapted to spatial homogeneity it can be shown 
\cite{Apostolopoulos:2003uu, Apostolopoulos:2004fd} that the HVF assumes the
form $\mathbf{H}=\psi t\partial _{t}+H^{\alpha }(x^{\beta })\partial
_{\alpha }$.

\noindent On pure geometrical grounds, the existence of a HVF \cite%
{Hall-Book} 
\begin{eqnarray}
R_{abcd}H^{d} &=&F_{ab;c},\hspace{0.4cm}\mathcal{L}_{\mathbf{H}}\Gamma _{%
\hspace{0.1cm}cd}^{a}=0,\hspace{0.4cm}\mathcal{L}_{\mathbf{H}}R_{\hspace{%
0.15cm}bcd}^{a}=0  \nonumber \\
&&  \nonumber \\
\mathcal{L}_{\mathbf{H}}R_{ab} &=&0,\hspace{0.4cm}\mathcal{L}_{\mathbf{H}%
}R=-2\psi R,\hspace{0.4cm}\mathcal{L}_{\mathbf{H}}G_{ab}=0
\label{HomothetRel1}
\end{eqnarray}%
where $F_{ab}\equiv H_{[a;b]}$ is the homothetic bivector and $G_{ab}=R_{ab}-%
\frac{R}{2}g_{ab}$ is the Einstein tensor. Equations (\ref{HVFEquation1})
and (\ref{HomothetRel1}) mean that for a \emph{proper} HVF (i.e. $\psi \neq
0 $ and $\psi _{;a}=0$) the geometrical quantities of the SH models scale $%
\sim t^{p}$. In particular the curvature scalar $R$ satisfies 
\begin{equation}
R=R_{0}t^{-2}.  \label{CurvatureScalar1}
\end{equation}%
It should be also noticed that the invariance of the connection coefficients
along the integral curves of a homothetic symmetry implies that the Lie and
covariant derivatives commute \cite{Yano} 
\begin{equation}
\mathcal{L}_{\mathbf{H}}\nabla _{a}\Phi =\nabla _{a}\mathcal{L}_{\mathbf{H}%
}\Phi  \label{Commut1}
\end{equation}%
for any scalar or tensorial quantity $\Phi $.

\noindent Similar transformation mappings on the kinematical quantities of
the unit timelike vector field $u^{a}$ do not necessarly hold unless the
homothetic symmetry \textit{is inherited} by $u^{a}$ i.e. $\mathcal{L}_{%
\mathbf{H}}u^{a}=-\psi u^{a}\Leftrightarrow \mathcal{L}_{\mathbf{H}%
}u_{a}=\psi u_{a}$. Although, in general, this is a consequence of the full
FEs in the case of SH models it can be shown \cite{Apostolopoulos:2003uu,
Apostolopoulos:2004fd} that the inheritance property is an intrinsic feature
of its geometric structure. As a result and using equation (\ref{Commut1})
we get 
\begin{equation}
\mathcal{L}_{\mathbf{H}}u_{a;b}=\psi u_{a;b},\hspace{0.4cm}\mathcal{L}_{%
\mathbf{H}}\sigma _{ab}=\psi \sigma _{ab},\hspace{0.4cm}\mathcal{L}_{\mathbf{%
H}}\theta =-\psi \theta .  \label{KinematInheritance1}
\end{equation}%
i.e. the kinematical quantities of the timelike congruence scale as $t^{-1}$
(e.g. the expansion rate has the form $\theta =\theta _{0}t^{-1}$).

\noindent The projection tensor also inherits the homothetic symmetry 
\begin{equation}
\mathcal{L}_{\mathbf{H}}h_{ab}=2\psi h_{ab}
\label{LieDerivativeProjectionTensor1}
\end{equation}%
which yields to the following useful identity: 
\begin{equation}
\mathcal{L}_{\mathbf{H}}\left( h_{a}^{c}h_{b}^{d}-\frac{1}{3}%
h^{cd}h_{ab}\right) =0  \label{LieDerivativeProjectionTensor2}
\end{equation}%
where the $h_{a}^{c}h_{b}^{d}-\frac{1}{3}h^{cd}h_{ab}$ operator returns the $%
u^{a}-$normal and trace-free part of any second order symmetric tensor.

\noindent The above assumptions can be seen as purely geometrical therefore
they are valid irrespective of the form of the FEs. One should expect that
merging the dynamics of the system, embodied in the energy-momentum (EM)
tensor $T_{ab}$, with the geometry of the SH\ models further restrictions in
both sectors will appear. The standard way to analyze the dynamical content
of any model is to decompose the associated EM tensor $T_{ab}$ into
irreducible parts w.r.t. $u^{a}$:

\begin{equation}
T_{ab}=\rho u_{a}u_{b}+ph_{ab}+q_{(a}u_{b)}+\pi _{ab}
\label{TensorDecomposition1}
\end{equation}
where

\begin{eqnarray}
\rho &\equiv &T_{ab}u^{a}u^{b}\hspace{0.4cm}p\equiv \frac{1}{3}T_{ab}h^{ab} 
\nonumber \\
&&  \nonumber \\
q_{a} &\equiv &-h_{a}^{c}T_{cd}u^{d}\hspace{0.4cm}\pi _{ab}\equiv \left(
h_{a}^{c}h_{b}^{d}-\frac{1}{3}h^{cd}h_{ab}\right) T_{cd}.
\label{IrreducParts}
\end{eqnarray}%
In any matter fluid model $\rho $, $p$, $q_{a}$ and $\pi _{ab}$ represent
the energy density, the isotropic pressure, the heat flux vector field and
the anisotropic pressure tensor respectively obeying the standard energy
conditions \cite{Wald}. Nevertheless the decomposition (\ref%
{TensorDecomposition1}) remains true for every symmetric second order tensor
and, as we shall see in the next section, will be used to set up the system
of equations for the vacuum $F(R)$-gravity in terms of dimensionless
variables.

\noindent In order to visualize how the homothetic symmetry interacts with
the dynamics, it is necessary to determine the effect of the former on the
irreducible parts. Provided that $\mathcal{L}_{\mathbf{H}}u^{a}=-\psi u^{a}$
and $\mathcal{L}_{\mathbf{H}}T_{ab}=0$ (which is the direct consequence of (%
\ref{HomothetRel1}) and the FEs $G_{ab}=T_{ab}$), the quantities (\ref%
{IrreducParts}) are Lie transformed along $H^{a}$ according to 
\begin{eqnarray}
\mathcal{L}_{\mathbf{H}}\rho &=&-2\psi \rho ,\hspace{0.4cm}\mathcal{L}_{%
\mathbf{H}}p=-2\psi p  \nonumber \\
&&  \nonumber \\
\mathcal{L}_{\mathbf{H}}q_{a} &=&-2\psi q_{a},\hspace{0.4cm}\mathcal{L}_{%
\mathbf{H}}\pi _{ab}=0  \label{InheritanceIrreducible}
\end{eqnarray}%
as we can verify using equations (\ref{KinematInheritance1})-(\ref%
{LieDerivativeProjectionTensor2}).

\noindent The first two equations imply that the equation of state of a
self-similar fluid model is necessarily linear i.e. $p=w\rho $ where $w$ is
the (constant) state parameter. In the particular case of SH models $\rho
=\rho _{0}t^{-2}$ and $p=p_{0}t^{-2}$.

\section{Spatially homogeneous vacuum cosmologies in $F(R)-$ gravity}

\setcounter{equation}{0}

\noindent The effective action we are interested has the form 
\begin{equation}
S=\int d^{4}x\sqrt{-g}\,\left[ \mathcal{L}^{\mathrm{mat}}+\,F(R)\right]
\label{action1}
\end{equation}%
where $F(R)$ is an analytic function of the curvature scalar $R$.

\noindent Considering only the metric as the independent variable, the
extrema of the action (\ref{action1}) give the effective FEs in the form 
\cite{Sotiriou:2008rp, DeFelice:2010aj} 
\begin{equation}
\Psi _{ab}\equiv F_{,R}(R)R_{ab}-\frac{1}{2}F(R)g_{ab}-\nabla _{a}\nabla
_{b}F_{,R}(R)+g_{ab}\nabla ^{2}F_{,R}(R)=T_{ab}^{\mathrm{mat}}
\label{FieldEquations1}
\end{equation}%
where $F_{,R}(R)=\frac{dF}{dR}$ and $T_{ab}^{\mathrm{mat}}$ is the
energy-momentum (EM) tensor representing the matter contributions in the
dynamics of the model from a continuum system. It is easy to see that the
tensor $\Psi _{ab}$ is divergence-free which implies the usual energy and
momentum conservation. Taking the trace of (\ref{FieldEquations1}) 
\begin{equation}
\nabla ^{2}F_{,R}=\frac{1}{3}\left( 2F-F_{,R}R+T^{\mathrm{mat}}\right)
\label{TraceFE1}
\end{equation}%
and substituting back to (\ref{FieldEquations1}) we get 
\begin{equation}
F_{,R}R_{ab}+\frac{1}{6}g_{ab}\left( F-2F_{,R}R+2T^{\mathrm{mat}}\right)
-\nabla _{a}\nabla _{b}F_{,R}=T_{ab}^{\mathrm{mat}}.  \label{FieldEquations2}
\end{equation}%
The differential equation (\ref{TraceFE1}) imposes further constraints on
the geometric structure of the model by restricting the functional form of $%
F(R)$. This will become more transparent when we will include the
self-similarity property.

\noindent We confine our study to SH vacuum models i.e. $T_{ab}^{\mathrm{mat}%
}=0$ and observe that, formally, the effective FEs (\ref{FieldEquations1})
can be written in the familiar form of standard gravity as (provided that $%
F_{,R}\neq 0$) 
\begin{equation}
G_{ab}=T_{ab}^{\mathrm{eff}}  \label{FieldEquations3}
\end{equation}%
where 
\begin{equation}
T_{ab}^{\mathrm{eff}}\equiv -\frac{1}{6}g_{ab}\left( \frac{F}{F_{,R}}%
+R\right) +\frac{1}{F_{,R}}\nabla _{a}\nabla _{b}F_{,R}.
\label{EffectiveEnergyMomentum1}
\end{equation}%
Because the curvature scalar is also SH, the tensor $\nabla _{a}\nabla
_{b}F_{,R}$ can be written in terms of the kinematical quantities (\ref%
{Veloc-Parts}) of the timelike vector field $u^{a}$ 
\begin{equation}
\nabla _{a}\nabla _{b}F_{,R}=\left( F_{,R}\right) ^{\cdot \cdot
}u_{a}u_{b}-\left( F_{,R}\right) ^{\cdot }\left( \sigma _{ab}+\frac{\theta }{%
3}h_{ab}\right)  \label{SecondDerivatives1}
\end{equation}%
where a dot \textquotedblleft $\cdot $\textquotedblright\ denotes
differentiation w.r.t. $u^{a}$ (equivalently the time coordinate).

\noindent Taking the trace of (\ref{SecondDerivatives1}) and using (\ref%
{TraceFE1}) we get:

\begin{equation}
\nabla _{a}\nabla _{b}F_{,R}=\left[ \frac{1}{3}\left( F_{,R}R-2F\right)
-\left( F_{,R}\right) ^{\cdot }\theta \right] u_{a}u_{b}-\left(
F_{,R}\right) ^{\cdot }\left( \sigma _{ab}+\frac{\theta }{3}h_{ab}\right)
\label{SecondDerivatives2}
\end{equation}%
Equations (\ref{IrreducParts}) and (\ref{FieldEquations3}) in conjuction
with (\ref{SecondDerivatives2}) imply that $F(R)$ vacuum models with an
underlying SH geometry can be seen as \textit{anisotropic} \textquotedblleft
fluid\textquotedblright\ models with \textquotedblleft
dynamical\textquotedblright\ quantities satisfying

\begin{equation}
\rho =\frac{1}{2}\left( R-\frac{F}{F_{,R}}\right) -\frac{\left(
F_{,R}\right) ^{\cdot }}{F_{,R}}\theta  \label{EnergyDensity1}
\end{equation}%
\begin{equation}
p=-\frac{1}{6}\left( R+\frac{F}{F_{,R}}\right) -\frac{1}{3}\frac{\left(
F_{,R}\right) ^{\cdot }}{F_{,R}}\theta  \label{IsotropicPressure1}
\end{equation}%
\begin{eqnarray}
q_{a} &=&-T_{cd}^{\mathrm{eff}}u^{c}h_{a}^{d}=0  \nonumber \\
&&  \nonumber \\
\pi _{ab} &=&\left( h_{a}^{c}h_{b}^{d}-\frac{1}{3}h^{cd}h_{ab}\right)
T_{cd}^{\mathrm{eff}}=-\frac{\left( F_{,R}\right) ^{\cdot }}{F_{,R}}\sigma
_{ab}.  \label{StressesSH1}
\end{eqnarray}%
As a result every SH vacuum model in $F(R)-$gravity is equivalent with a
\textquotedblleft flux\textquotedblright -free anisotropic fluid model in
standard gravity where the \textquotedblleft dissipative\textquotedblright\
tensor satisfies the Eckart-Landau-Lifshitz relation \cite{Leon:2010pu}.

\noindent In conventional fluid models of standard gravity there are no
evolution equations for the isotropic pressure $p$ and the anisotropic
stress tensor $\pi _{ab}$. Therefore equations (\ref{TraceFE1}), (\ref%
{SecondDerivatives1}), (\ref{SecondDerivatives2}) and (\ref{EnergyDensity1}%
)-(\ref{StressesSH1}) \textit{completely} determine the dynamics of the
vacuum SH models in $F(R)-$gravity provided that the functional form of $%
F(R) $ is known. One can then exploit the orthonormal frame formalism and
the usage of expansion-normalized variables \cite{Hewitt:2000nx} to
reformulate the FEs (\ref{FieldEquations3}) as evolution equations of the
shear and spatial curvature (described by the variables $A_{\alpha }$ and $%
N_{\alpha \beta }$ which identify each Bianchi type model). The evolution
equations are subjected to algebraic constraints and can be used to study
the intermediate and asymptotic behaviour of general SH\ models in $F(R)-$%
gravity. We postpone this analysis for a future work. In the next section we
use the set of equations of \cite{Hewitt:2000nx} as a guide in order to
examine the existence of a particular self-similar model in $F(R)$ vacuum
gravity and to give the corresponding exact solution whenever it exists.

\section{The implications of self-similarity}

\setcounter{equation}{0}

\noindent The \textquotedblleft flux\textquotedblright -free property and
the presence of an additional contribution in the evolution equation of the
shear as well as the \emph{algebraic constraint} (\ref{EnergyDensity1}) are
intrinsic properties of the vacuum models in $F(R)-$ gravity that allow us
to close the system of equations. On the other hand the identification of
self similar models as equilibrium points in the dynamical phase space of
general configurations shows that the existence or not of a proper HVF is
crucial in understanding the structural form of the associated state space.
Nevertheless we expect that the assumption of self-similarity will provide
us with further, \emph{algebraic} in nature, restrictions.

\noindent From the inheritance property of the shear (\ref%
{KinematInheritance1}), the invariance under Lie dragging of the anisotropic
stress tensor (\ref{InheritanceIrreducible}) and equation (\ref{StressesSH1}%
) imply that 
\begin{equation}
\mathcal{L}_{\mathbf{H}}\left[ \frac{\left( F_{,R}\right) ^{\cdot }}{F_{,R}}%
\right] =-\psi \frac{\left( F_{,R}\right) ^{\cdot }}{F_{,R}}.
\label{EvolutionF(R)}
\end{equation}%
Expressing the last equation in coordinate form we can easily verify that
the functional $F(R)$ satisfies 
\begin{equation}
F(R)=F_{0}R^{n}  \label{FunctionalForm1}
\end{equation}%
where $F_{0}$ is an arbitrary constant and $n$ is any real number.

\noindent Clearly, the power law behaviour of $F(R)$ is a \emph{direct
consequence of the scale invariant feature (self-similarity) of the SH
vacuum models in modified gravity}. We note however that in evolving models
the functional structure of $F(R)$ is more general and complicated. The
established power law property of $F(R)$ permit us to determine the exact
form of the \textquotedblleft anisotropic\textquotedblright\ stress tensor $%
\pi _{ab}$. From equation (\ref{StressesSH1}) a straightforward calculation
gives%
\begin{equation}
\pi _{ab}=\frac{6\left( n-1\right) }{\theta _{0}}H\sigma _{ab}\equiv \tilde{A%
}H\sigma _{ab}.  \label{AnisotropicPressure1}
\end{equation}%
We have employed the Hubble scalar defined as 
\begin{equation}
H\equiv \frac{\theta }{3}=\frac{\theta _{0}}{3t}.  \label{HubbleScalar1}
\end{equation}%
Although we have used the standard form for the FEs (\ref{FieldEquations3})
we must emphasize that the definition of the effective \textquotedblleft
energy\textquotedblright\ density (\ref{EnergyDensity1}) or the effective
isotropic \textquotedblleft pressure\textquotedblright\ (\ref%
{IsotropicPressure1}) reveal further constraints which one must take into
account. In the particular case of self-similar models the induced linear
dependence $p=w\rho $ implies that only one of them contains non trivial
information and the other is satisfied identically. It follows from eqs (\ref%
{FunctionalForm1})-(\ref{AnisotropicPressure1}) in conjuction with the trace 
$R=\rho -3p=\rho (1-3w)$ of the effective FEs that the \textquotedblleft
energy\textquotedblright\ density satisfies the algebraic relation 
\begin{equation}
\rho =\rho \frac{1-3w}{2}\frac{n-1}{n}+3H^{2}\tilde{A}.
\label{EnergyConstraint1}
\end{equation}%
We observe that for $n=1$ or equivalently $\tilde{A}=0$ we reproduce the
usual vacuum models in standard gravity ($\tilde{A}=0\Leftrightarrow \pi
_{ab}=0$, $\rho =p=0$).

\subsection{\noindent The FEs in expansion normalized variables}

\noindent Of particular importance in the exploration of the asymptotic
dynamics of SH models, is the reformulation of the complete set of the FEs (%
\ref{FieldEquations3}) as autonomous system of first order ordinary
differential equations. This can be done by defining a set of
expansion-normalized (dimensionless) variables 

\begin{equation}
\Sigma _{\alpha \beta }=\frac{\sigma _{\alpha \beta }}{H},\qquad N_{\alpha
\beta }=\frac{n_{\alpha \beta }}{H},\qquad A_{\alpha }=\frac{a_{\alpha }}{H}
\label{expansion-normalized1}
\end{equation}%
\begin{equation}
R_{\alpha }=\frac{\Omega _{\alpha }}{H},\qquad \Omega =\frac{\rho }{3H^{2}}%
,\qquad P=\frac{p}{3H^{2}}  \label{expansion-normalized2}
\end{equation}%
\begin{equation}
Q_{\alpha }=\frac{q_{\alpha }}{H^{2}},\qquad \Pi _{\alpha \beta }=\frac{\pi
_{\alpha \beta }}{H^{2}}.  \label{expansion-normalized3}
\end{equation}%
where the greek indices reflect the orthonormal frame $\{\mathbf{\omega }%
^{\alpha }\}$ componets of the kinematical variables ($\sigma _{\alpha \beta
}$), the rate of rotation of the spatial frame ($\Omega _{\alpha }$), the
spatial rotation (commutators of the dual $\mathbf{e}_{\alpha }(t)$ of the
1-forms $\mathbf{\omega }^{\alpha }$) variables ($a_{\alpha }$, $n_{\alpha
\beta }$) and the dynamical quantities ($\rho $, $p$, $q_{\alpha }$, $\pi
_{\alpha \beta }$) \cite{Ellis:1998ct}. It is convenient to define the shear
parameter $\Sigma $ 
\begin{equation}
\Sigma ^{2}=\frac{\sigma ^{2}}{3H^{2}}.  \label{SigmaSquare1}
\end{equation}%
The (dimensionless) spatial curvature variables have the form 
\[
S_{\alpha \beta }=\frac{^{3}S_{\alpha \beta }}{H^{2}},\qquad K=-\frac{^{3}R}{%
6H^{2}}
\]%
where $^{3}S_{\alpha \beta }$ and $^{3}R$ are the trace-free and the trace
of the Ricci tensor of the instantaneous rest space of the comoving
observers $u^{a}$. From the first of equations (\ref{expansion-normalized1}) and
eq. (\ref{SigmaSquare1}) it follows 
\begin{equation}
\Sigma ^{2}=\frac{1}{6}\Sigma _{\alpha \beta }\Sigma ^{\alpha \beta }.
\label{SigmaSquare2}
\end{equation}%
It can be shown (equations (1.69)-(1.70) of \cite{Wainwright-Ellis}) that 
\[
S_{\alpha \beta }=B_{\alpha \beta }-\frac{1}{3}B_{\mu }^{~\mu }\delta
_{\alpha \beta }-2\varepsilon ^{\mu \nu }{}_{(\alpha }N_{\beta )\mu }A_{\nu
},\qquad K=\frac{1}{12}B_{\mu }^{~\mu }+A_{\mu }A^{\mu }.
\]%
where 
\[
B_{\alpha \beta }=2N_{\alpha }^{~\mu }N_{\mu \beta }-N_{\mu }^{~\mu
}N_{\alpha \beta }.
\]%
The key ingredient necessary to construct the associate dynamical system
that follows from the FEs (\ref{FieldEquations3}) is to define the
dimensionless time variable $\tau $ according to 
\begin{equation}
\frac{dt}{d\tau }=\frac{1}{H},\quad \frac{dH}{d\tau }=-\left( 1+q\right) H
\label{diffequat2}
\end{equation}%
where $q$ is the deceleration parameter and $H$ is the Hubble scalar. This
results the decoupling of the evolution equation of $H=\theta /3$ from the
rest of the evolution equations (eqs (1.90)-(1.100) of \cite%
{Wainwright-Ellis}) and the full set becomes \cite%
{Hewitt:2000nx} (a prime \textquotedblleft $^{\prime }$\textquotedblright\
denotes differentiation w.r.t. $\tau $) 

\begin{equation}
\Sigma _{\alpha \beta }^{\prime }=-\left( 2-q\right) \Sigma _{\alpha \beta
}+2\epsilon _{\hspace{0.3cm}(\alpha }^{\mu \nu }\Sigma _{\beta )\mu }R_{\nu
}-S_{\alpha \beta }+\tilde{A}\Sigma _{\alpha \beta }  \label{evol1}
\end{equation}%
\begin{equation}
N_{\alpha \beta }^{\prime }=qN_{\alpha \beta }+2\Sigma _{(\alpha }^{\hspace{%
0.2cm}\mu }N_{\beta )\mu }+2\epsilon _{\hspace{0.3cm}(\alpha }^{\mu \nu
}N_{\beta )\mu }R_{\nu }  \label{evol2}
\end{equation}%
\begin{equation}
A_{\alpha }^{\prime }=qA_{\alpha }-\Sigma _{\alpha }^{\hspace{0.2cm}\mu
}A_{\mu }+\epsilon _{\alpha }^{\hspace{0.2cm}\mu \nu }A_{\mu }R_{\nu }
\label{evol3}
\end{equation}%
\begin{equation}
\Omega ^{\prime }=(2q-1-3w)\Omega -\frac{1}{3}\tilde{A}\Sigma _{\mu \nu
}\Sigma ^{\mu \nu }  \label{Bianchi1}
\end{equation}%
\begin{equation}
Q_{\alpha }^{\prime }=3\tilde{A}A^{\beta }\Sigma _{\alpha \beta }+\epsilon
_{\alpha }^{\hspace{0.2cm}\mu \nu }N_{\mu }^{\hspace{0.1cm}\beta }\Sigma
_{\beta \nu }  \label{EvolutionFlux1}
\end{equation}%
\begin{equation}
N_{\alpha }^{\hspace{0.1cm}\beta }A_{\beta }=0  \label{Jacobi1}
\end{equation}%
\begin{equation}
\Omega =1-\Sigma ^{2}-K  \label{Friedman1}
\end{equation}

\begin{equation}
3A^{\beta }\Sigma _{\alpha \beta }-\epsilon _{\alpha }^{\hspace{0.2cm}\mu
\nu }N_{\mu }^{\hspace{0.1cm}\beta }\Sigma _{\beta \nu }=0
\label{algebraic2}
\end{equation}%
\begin{equation}
\Omega =\Omega \frac{1-3w}{2}\frac{n-1}{n}+\tilde{A}
\label{EnergyCostraint2}
\end{equation}%
where for the rhs we have used eqs. (\ref{AnisotropicPressure1})-(\ref%
{EnergyConstraint1}). 

The above non linear system of first-order differential equations is
sufficient to locate any equilibria (i.e. when $\Sigma _{\alpha \beta
}^{\prime }=N_{\alpha \beta }^{\prime }=0=\Omega ^{\prime }$ and $Q_{\alpha
}^{\prime }=0=Q_{\alpha }$) of the dynamical system in vacuum $F(R)-$%
gravity. However, two more restrictions exist and can be used as
auxiliary relations due to their simple form. The first follows from (\ref%
{HubbleScalar1}) and (\ref{diffequat2}) namely 
\begin{equation}
\theta _{0}(1+q)=3  \label{DecelExpans1}
\end{equation}%
and the second is the definition of the dimensionless constant $\tilde{A}$
which, in some sense, represents the deviation from standard gravity 
\begin{equation}
\tilde{A}=\frac{6\left( n-1\right) }{\theta _{0}}.  \label{DefinitionA}
\end{equation}%
The remaining freedom of a time-dependent spatial rotation permit us to
choose the orthonormal tetrad to be the eigenframe of $N_{\alpha \beta }$
therefore the contracted form of Jacobi identities $N_{\alpha \beta
}A^{\beta }=0$ implies:

\begin{equation}
N_{\alpha \beta }=\left( 
\begin{array}{lll}
N_{1} & 0 & 0 \\ 
0 & N_{2} & 0 \\ 
0 & 0 & N_{3}%
\end{array}%
\right) ,\qquad A_{\alpha }=A_{1}\delta _{\alpha }^{1}  \label{def1}
\end{equation}%
where the value of $A_{\alpha }$ distinguishes the models in class A ($A_1=0$%
) and class B ($A_{1}\neq 0$).

\noindent In this case the angular velocity $R_{\alpha }=[R_{1},R_{2},R_{3}]$
of the spatial orthonormal frame will be specified from (\ref{evol1})-(\ref%
{DefinitionA})\ as function of the shear variables. Furthermore the
differential \textquotedblleft versions\textquotedblright\ of (\ref{evol2})
and (\ref{evol3}) have a first integral and the component $A_{1}$ is
expressed in the well known form: 
\begin{equation}
A_{1}^{2}=hN_{2}N_{3}.  \label{first_integral}
\end{equation}%
Especially in type $VI_{h}$ (where $N_{2}N_{3}<0$) the distinction between
models with $h\neq -1/9$ and $h=-1/9$ is the reminiscence of the \emph{%
exceptional} algebraic behaviour of the $0\alpha -$components of the FEs.

\noindent Throughout the rest of the present section we found some new \emph{%
self similar SH vacuum} and \emph{anisotropic} ($\sigma _{\alpha \beta }\neq
0$) models in \emph{proper} $F(R)-$gravity ($R\neq 0$) and give their local
metric form using the results of \cite{Apostolopoulos:2003uu,
Apostolopoulos:2004fd}. It should be noticed that the equilibria of the
state space which correspond to vanishing shear (Robertson-Walker geometry)
or the case where the curvature scalar is zero \footnote{%
We note that the Bianchi type I solutions found in \cite{Barrow:2005dn,
Clifton:2006kc} correspond either to a RW geometry or to a model with $R=0$
(see also \cite{Leach:2006br})} (in our notation the constraint $R=0$ is
equivalent with a radiation-like fluid where the state parameter satisfies $%
w=1/3$) are not of lesser importance to the qualitative study of SH vacuum
models and will be reported in forthcoming works together with a detailed
analysis of their asymptotical and intermediate behaviour.\newline

\noindent \emph{\textbf{SH models of Bianchi type II}}

\noindent Substituting $A_{\alpha }=0$ and $N_{2}=N_{3}=0$ back to (\ref%
{evol1})-(\ref{DefinitionA}) it follows that $R_{\alpha }=0=\Sigma _{\alpha
\beta }$ ($\alpha \neq \beta $) i.e. the model is \textquotedblleft
diagonal\textquotedblright\ (the off-diagonal shear components vanishes).
Solving the remaining equations we found the following fixed point 
\begin{equation}
N_{1}=\frac{6\sqrt{-\left( 2n^{2}-2n-1\right) \left( 11n^{2}-20n+8\right) }}{%
16n^{2}-25n+10}  \label{Curvature_Variable_Type_II}
\end{equation}%
\begin{equation}
\Sigma _{22}=\Sigma _{33}=-\frac{2\left( 2n^{2}-2n-1\right) }{16n^{2}-25n+10}%
.  \label{Shear_Variable_Type_II}
\end{equation}%
\begin{equation}
q=4\Sigma _{22}.  \label{Deceleration_II}
\end{equation}%
The effective \textquotedblleft energy\textquotedblright\ density and the
state parameter are%
\begin{equation}
\Omega =\frac{18\left( n-1\right) \left( 17n^{3}-36n^{2}+24n-4\right) }{%
\left( 16n^{2}-25n+10\right) ^{2}}  \label{EffectiveEnergy_II}
\end{equation}%
\begin{equation}
w=-\frac{49n^{3}-84n^{2}+24n+4}{3\left( 17n^{3}-36n^{2}+24n-4\right) }.
\label{StateParameter_II}
\end{equation}%
It is worth noticing that we do not require the positivity of $\Omega $
since the effective EM tensor does not represent an actual matter fluid.
Nevertheless the ``energy'' condition $\Omega >0$ could be imposed in
constructing a compact phase space for each model \cite{Goheer:2007wu}.

\noindent The above new exact anisotropic model of Bianchi type II in vacuum 
$F(R)-$gravity admits the HVF 
\begin{equation}
\mathbf{H}=\frac{3\left( n-2\right) }{2\left( 2n^{2}-2n-1\right) }t\mathbf{%
\partial }_{t}+x\mathbf{\partial }_{x}+2y\mathbf{\partial }_{y}+z\mathbf{%
\partial }_{z}  \label{HVF_Bianci_II}
\end{equation}%
and the line element assumes the form 
\begin{eqnarray}
ds^{2} &=&-dt^{2}+\frac{1}{2}%
t^{2p_{1}}dx^{2}+D(n)t^{2p_{2}}dy^{2}-2D(n)xt^{2p_{2}}dydz+  \nonumber \\
&&  \nonumber \\
&&+\left[ D(n)x^{2}t^{2p_{2}}-\frac{1}{2}t^{2p_{1}}\right] dz^{2}.
\label{Bianchi_II_Exact}
\end{eqnarray}%
The $p_{1},p_{2}$ indices and the polyonym $D(n)$ satisfy the relations 
\begin{equation}
p_{1}=\frac{4n^{2}-7n+4}{3\left( 2-n\right) },\qquad p_{2}=\frac{8n^{2}-11n+2%
}{3\left( 2-n\right) }  \label{Bianchi_II_Indices}
\end{equation}%
\begin{equation}
D(n)=\frac{22n^{4}-62n^{3}+45n^{2}+4n-8}{9\left( n-2\right) ^{2}}.
\label{Polyonym_Bianchi_II}
\end{equation}%
The negativity of the det$(g)$ or equivalently the real values of the
curvature variable $N_{1}$ imply 
\begin{equation}
n\in \left( \frac{1}{2}-\frac{\sqrt{3}}{2},\frac{10}{11}-\frac{2\sqrt{3}}{11}%
\right) \vee \left( \frac{10}{11}+\frac{2\sqrt{3}}{11},\frac{1}{2}+\frac{%
\sqrt{3}}{2}\right) .  \label{Range_Bianchi_II}
\end{equation}%
\bigskip \emph{\textbf{SH models of Bianchi type VI}}$_{0}$

\noindent Similarily with the previous type we get $R_{\alpha }=0=\Sigma
_{\alpha \beta }$ ($\alpha \neq \beta $) and the curvature variables satisfy 
$N_{2}+N_{3}=0$ therefore the fixed point corresponds to the subclass $N_{%
\hspace{0.2cm}\alpha }^{\alpha }=0$. The kinematical and dynamical variables
of this model are given by 
\begin{equation}
N_{2}=-\frac{3\sqrt{-3\left( 2n^{2}-2n-1\right) }\left\vert n-1\right\vert }{%
4n^{2}-7n+4}  \label{Curvature_Variable_Type_VI0}
\end{equation}%
\begin{equation}
\Sigma _{22}=\Sigma _{33}=\frac{2n^{2}-2n-1}{4n^{2}-7n+4}
\label{Shear_Variable_Type_VI0}
\end{equation}%
\begin{equation}
\Omega =\frac{6\left( n-1\right) \left( 5n^{3}-12n^{2}+9n-1\right) }{\left(
4n^{2}-7n+4\right) ^{2}}  \label{EffectiveEnergy_VI0}
\end{equation}%
\begin{equation}
w=-\frac{13n^{3}-24n^{2}+9n+1}{3\left( 5n^{3}-12n^{2}+9n-1\right) }.
\label{StateParameter_VI0}
\end{equation}%
\begin{equation}
q=-2\Sigma _{22}.  \label{Deceleration_VI0}
\end{equation}%
The generator of the self-similarity and the metric are 
\begin{equation}
\mathbf{H}=\frac{n-2}{2n^{2}-2n-1}t\mathbf{\partial }_{t}+\mathbf{\partial }%
_{x}+2y\mathbf{\partial }_{y}  \label{HVF_Bianchi_VI0}
\end{equation}

\begin{equation}
ds^{2}=-dt^{2}+t^{2}D(n)dx^{2}+t^{2p_{1}}e^{-2x}dy^{2}+t^{2p_{1}}e^{2x}dz^{2}
\label{BianchiVI0Exact}
\end{equation}%
where 
\begin{equation}
p_{1}=\frac{2n^{2}-3n+1}{2-n}  \label{Bianchi_VI0_Index}
\end{equation}%
\begin{equation}
D(n)=-\frac{\left( n-2\right) ^{2}}{3\left( n-1\right) ^{2}\left(
2n^{2}-2n-1\right) }.  \label{Polyonym_Bianchi_VI0}
\end{equation}%
This solution is well defined within the range

\begin{equation}
n\in \left( \frac{1}{2}-\frac{\sqrt{3}}{2},\frac{1}{2}+\frac{\sqrt{3}}{2}%
\right) .  \label{Range_Bianchi_VI0}
\end{equation}%
\bigskip \emph{\textbf{SH models of Bianchi type VI}}$_{h}$ $(h\neq -1/9)$

\noindent In comparison with the standard gravity vacuum anisotropic
cosmologies, the type $VI_{h}$ models exhibit common features and
significant differences as well. In particular we found that a solution
exists only within the subclass satisfying $N_{\hspace{0.2cm}\alpha
}^{\alpha }=0$, similar to the case $n=1$, namely 
\begin{equation}
N_{2}=-\frac{3\sqrt{\left[ h^{2}\left( n-2\right) ^{2}+3\left( n-1\right)
^{2}\right] \left( -2n^{2}+2n+1\right) }}{\left\vert 3h^{2}\left( n-2\right)
-4n^{2}+7n-4\right\vert }  \label{Curvature_Variable_Type_VIh}
\end{equation}%
\begin{equation}
\Sigma _{22}=\Sigma _{33}=-\frac{2n^{2}-2n-1}{3h^{2}\left( n-2\right)
-4n^{2}+7n-4}.  \label{Shear_Variable_Type_VIh}
\end{equation}%
\begin{equation}
\Omega =\frac{6\left( h^{2}+1\right) \left( n-1\right) \left[ 3h^{2}n\left(
n-2\right) ^{2}+5n^{3}-12n^{2}+9n-1\right] }{\left[ 3h^{2}\left( n-2\right)
-4n^{2}+7n-4\right] ^{2}}  \label{EffectiveEnergy_VIh}
\end{equation}%
\begin{equation}
w=-\frac{3h^{2}n\left( n-2\right) ^{2}+13n^{3}-24n^{2}+9n+1}{3\left[
3h^{2}n\left( n-2\right) ^{2}+5n^{3}-12n^{2}+9n-1\right] }.
\label{StateParameter_VIh}
\end{equation}%
\begin{equation}
q=-2\Sigma _{22}.  \label{Deceleration_VIh}
\end{equation}%
From (\ref{Curvature_Variable_Type_VIh}) we deduce that the range of
parameter $n$ is the same like the type VI$_{0}$.

\noindent The FEs (\ref{FieldEquations1}) and the usage of the associated
self-similar metric \cite{Apostolopoulos:2004fd} yield

\begin{equation}
\mathbf{H}=\frac{\left( 2-n\right) \left( h^{2}+1\right) }{\left(
2n^{2}-2n-1\right) \left( h-1\right) }t\mathbf{\partial }_{t}+\frac{h+1}{1-h}%
y\mathbf{\partial }_{y}+z\mathbf{\partial }_{z}  \label{HVF_Bianchi_VIh}
\end{equation}

\begin{equation}
ds^{2}=-dt^{2}+t^{2}D(n)dx^{2}+t^{2p_{1}}e^{-2x}dy^{2}+t^{2p_{2}}e^{2\frac{%
h+1}{h-1}x}dz^{2}  \label{BianchiVIhExact}
\end{equation}%
\begin{equation}
p_{1}=\frac{h^{2}\left( n-2\right) -h\left( 2n^{2}-2n-1\right) -2n^{2}+3n-1}{%
\left( n-2\right) \left( h^{2}+1\right) }  \label{Bianchi_VIh_Indices_1}
\end{equation}%
\begin{equation}
p_{2}=\frac{h^{2}\left( n-2\right) +h\left( 2n^{2}-2n-1\right) -2n^{2}+3n-1}{%
\left( n-2\right) \left( h^{2}+1\right) }  \label{Bianchi_VIh_Indices_2}
\end{equation}%
\begin{equation}
D(n)=-\frac{\left[ \left( h^{2}+1\right) \left( n-2\right) \right] ^{2}}{%
\left( h-1\right) ^{2}\left( 2n^{2}-2n-1\right) \left[ h^{2}\left(
n^{2}-4n+4\right) +3\left( n-1\right) ^{2}\right] }.
\label{Polyonym_Bianchi_VIh}
\end{equation}%
In contrast with the corresponding model in standard gravity \cite%
{Apostolopoulos:2004fd} the above solution does not admit a null gradient
KVF therefore its Petrov type is D. \bigskip

\noindent \emph{\textbf{SH models of the exceptional Bianchi type VI}}$%
_{-1/9}$

\noindent Again the only non trivial solution satisfies $N_{\hspace{0.2cm}%
\alpha }^{\alpha }=0$ and the parameters of the model are 
\begin{equation}
N_{2}=-\frac{3\sqrt{6\left( -4n^{2}+6n-1\right) }}{2\left\vert
2n+1\right\vert }  \label{Curvature_Variable_Type_VI-1/9}
\end{equation}%
\begin{equation}
\Sigma _{22}=\Sigma _{33}=\frac{n-2}{2n+1}=\Sigma _{23}
\label{Diagonal_Shear_Variables_Type_VI-1/9}
\end{equation}%
\begin{equation}
\Sigma _{12}=\Sigma _{13}=-\frac{\sqrt{10\left( -8n^{2}+14n-5\right) }}{%
2\left\vert 2n+1\right\vert }
\label{Non_Diagonal_Shear_Variables_Type_VI-1/9}
\end{equation}%
\begin{equation}
R_{1}=0,\qquad R_{2}=\Sigma _{13}=-R_{3}
\label{Angular_Velocity_Bianchi_VI-1/9}
\end{equation}%
\begin{equation}
\Omega =\frac{10\left( n-1\right) \left( 4n-1\right) }{\left( 2n+1\right)
^{2}}  \label{EffectiveEnergy_VIh_Exceptional}
\end{equation}%
\begin{equation}
w=\frac{1}{3\left( 1-4n\right) }.  \label{StateParameter_VIh_Exceptional}
\end{equation}%
\begin{equation}
q=-2\Sigma _{22}  \label{Deceleration_Exceptional}
\end{equation}%
The exact form of this exceptional anisotropic model is 
\begin{equation}
\mathbf{H}=\frac{1}{2-n}t\mathbf{\partial }_{t}+\frac{2}{5}y\mathbf{\partial 
}_{y}+\frac{4}{5}z\mathbf{\partial }_{z}  \label{HVF_Bianchi_VIh_Exceptional}
\end{equation}

\begin{eqnarray}
ds^{2}
&=&-dt^{2}+t^{2}D_{1}(n)dx^{2}+t^{2p_{1}}e^{-2x}dy^{2}+2e^{x/2}t^{2p_{1}}dxdz
\nonumber \\
&&  \nonumber \\
&&+t^{2p_{2}}D_{2}(n)dz^{2}  \label{Bianchi_Exceptional_Exact}
\end{eqnarray}%
\begin{equation}
p_{1}=\frac{2n+1}{5},\qquad p_{2}=\frac{4n-3}{5}
\label{Bianchi_Exceptional_Indices}
\end{equation}%
\begin{equation}
D_{1}(n)=\frac{75}{32\left( n^{2}-4n+4\right) }
\label{Polyonym_1_Bianchi_Exceptional}
\end{equation}%
\begin{equation}
D_{2}(n)=\frac{96\left( n^{2}-4n+4\right) \left( 4n^{2}-6n+1\right) }{%
125\left( 8n^{2}-14n+5\right) }.  \label{Polyonym_2_Bianchi_Exceptional}
\end{equation}%
It can be verified that this solution does not admit also a covariantly
constant null vector field which implies that the Petrov type is D (we
recall that $N_{\hspace{0.2cm}\alpha }^{\alpha }=0\Leftrightarrow \Sigma
_{22}=\Sigma _{33}$).\bigskip

\noindent \emph{\textbf{Non existence of self similar SH models}}

\noindent Regarding the rest of the Bianchi types our study showed that
no fixed points exist for the proper vacuum $F(R)-$gravity. For example
in type I the curvature variables are both zero ($A_{\alpha }=0=N_{\alpha
\beta }$) and the shear evolution equation implies $R_{\alpha }=0=\Sigma
_{\alpha \beta }$ ($\alpha \neq \beta $). It follows from the remaining set
of equations that \emph{they do not exist self-similar Bianchi type I anisotropic
vacuum models} as we can easily verify using also the FEs (\ref%
{FieldEquations1}) and the (diagonal) self similar three-dimensional metric
components $g_{\alpha \beta }=\mathtt{diag}(t^{p1},t^{p2},t^{p2})$.

\noindent This conclusion signifies a direct deviation from the
corresponding result of standard gravity where the Kasner circle plays a
crucial role in the dynamics of anisotropic models (either vacuum or non
vacuum) as past attractor \cite{Ellis:1998ct}. In addition the non-existence
of self-similar vacuum models of type III, IV and VII$_{h}$ contradicts our
expectations for a richer diversity of cosmological solutions due to the
fourth order of the resulting FEs and one must take into account additional
curvature invariants in the action integral \cite{Barrow:2006xb,
Middleton:2010bv}.

\section{Discussion}

\setcounter{equation}{0}

\noindent In this paper we found a set of new exact power-law solutions
which in principle can be used in order to understand the dynamics of
anisotropic vacuum or fluid models in power-law $F(R)-$gravity and analyze
their asymptotic behaviour \cite{Leon:2010pu, Leon:2013bra, Leon:2014dea}.
We emphasize that the analysis regarding the determination of the equilibria
was not exhausted since we have excluded various special cases e.g. the
shear- free models ($\sigma _{ab}=0$), the vanishing of the Ricci scalar
which represents a trivial solution of the FEs (\ref{FieldEquations1}) or
the stationary $q=-1$ cases. This does not mean that they are of minor
importance for the qualitative study of SH vacuum models. For example it has
been argued \cite{Leach:2006br} that the flat RW model in $F(R)-$gravity
could be a past attractor for anisotropic Bianchi type I models. Together
with the non existence of proper Kasner-like solutions this conclusion
implies that the curvature corrections dominate the shear influence and
opens the posibility for an isotropic singularity which is not occured in
standard gravity.

\noindent Another worth noticing point is the non existence of type III, IV
and VII$_{h}$ self-similar models in $F(R)-$gravity. A possible explanation
for this \textquotedblleft failure\textquotedblright\ is the fact that in
standard gravity the vacuum solutions e.g. of type III and IV are of Petrov
type N since they admit gradient null KVFs which essentially represent the
repeated principal null direction of the Weyl tensor and is the
characteristic feature of Kundt spacetimes \cite{Coley:2009ut}. A special
subset are the Vanishing Scalar Invariant (VSI) spacetimes where the type
III and IV solutions belong. Therefore we expect that including the self
similarity property to models of $F(R)-$gravity satisfying $R=0$,
algebraically special solutions could be found. On the other hand the
existence of type $VI_{h}$ solutions for any non-zero value of the group
parameter $h$ which are of Petrov type D (i.e. they are not VSI as their
counterparts in standard gravity) means that the effect of the curvature
corrections depends on the underlying geometric structure of the model.

\noindent It will be interesting to extent the efforts for the determination
of homothetic solutions in other categories of quadratic theories of gravity
where the Lagrangian permits the inclusion of curvature invariants like $%
R_{ab}R^{ab}$ or/and $R_{abcd}R^{abcd}$. Among other important issues, that
will show the level of departure from the well studied behaviour in standard
gravity and the role that self-similar models play in modified theories . We
believe that all these questions deserve further investigation and we intent
to study them in subsequent works keeping in mind that a more sophisticated
setup of the dynamical state space is required (see e.g. \cite{Alho:2016gzi}%
). \newline

\end{document}